\documentclass[12pt,preprint]{aastex}
\usepackage{epsfig}

\newcommand{\Hm}{\rm{H}^{-}}
\newcommand{\Hp}{\rm{H}^{+}}
\newcommand{\me}{\rm{e}}
\newcommand{\mH}{\rm{H}}
\newcommand{\mHt}{\rm{H}_{2}}
\newcommand{\mHtp}{\rm{H}_{2}^{+}}
\newcommand{\hi}{\hbox{H\,{\sc i}}\,} 

\begin{document}

\title{Comparing gas-phase and grain-catalyzed $\mHt$ formation}

\author{Simon C. O. Glover} 
\affil{Department of Astrophysics, American Museum of Natural History, \\
       Central Park West at 79th Street, New York, NY 10024}
\email{scog@amnh.org}

\begin{abstract}
Because $\mHt$ formation on dust grain surfaces completely dominates 
gas-phase $\mHt$ formation in local molecular clouds, it is often
assumed that gas-phase formation is never important. In fact, it is the 
dominant mechanism in a number of cases. In this paper, 
I briefly summarize the chemistry of gas-phase $\mHt$ formation,
and show that it dominates for dust-to-gas ratios less than a critical
value ${\cal D}_{\rm cr}$. I also show that ${\cal D}_{\rm cr}$ is simple
to calculate for any given astrophysical situation, and illustrate this 
with a number of examples, ranging from $\mHt$ formation in warm atomic gas
in the Milky Way to the formation of protogalaxies at high redshift.
\end{abstract}


\section{Introduction}
In local molecular clouds, molecular hydrogen ($\mHt$) forms primarily
on the surface of dust grains: two hydrogen atoms are adsorbed onto the 
surface of the grain and react to form $\mHt$, which subsequently escapes
back into the interstellar medium. However, $\mHt$ can also form in the
gas-phase, primarily through the reactions
\begin{eqnarray}
 \mH + \me & \rightarrow & \Hm + \gamma  \label{h2f1} \\
 \Hm + \mH & \rightarrow & \mHt + \me,   \label{h2f2}
\end{eqnarray}
although some also forms via the slower reactions
\begin{eqnarray}
 \mH + \Hp & \rightarrow & \mHtp + \gamma \label{h2f3} \\
 \mHtp + \mH & \rightarrow & \mHt + \Hp. \label{h2f4}
\end{eqnarray}
In dense gas, three-body reactions can also be important \citep{pss}, but 
these are ineffective at number densities $n < 10^{8} \: \rm{cm}^{-3}$.

Discussions of $\mHt$ formation have tended to concentrate on the role
played by dust, with little attention given to the gas-phase reactions.
However, as I show in section~\ref{astro}, in some circumstances these 
reactions can dominate the $\mHt$ formation rate.

In this paper, I briefly outline the chemistry of $\mHt$ formation and
show that it is easy to identify a critical dust-to-gas ratio 
${\cal D}_{\rm cr}$, above which grain-catalyzed formation dominates. 
I illustrate the method by applying it to various situations of astrophysical 
interest, and show that it can be a useful tool for estimating the 
importance of gas-phase $\mHt$ formation.

\section{The formation of molecular hydrogen}
\subsection{Gas-phase formation}
\label{gas}
Most of the molecular hydrogen that forms in the gas-phase does so via the
formation of an intermediate $\Hm$ ion, as outlined in reactions~\ref{h2f1} 
and \ref{h2f2} above. The first of these reactions occurs much more slowly than
the second, and so the equilibrium abundance of $\Hm$ is small and is rapidly 
reached. Thereafter, the $\mHt$ formation rate is determined by two factors: 
the rate at which $\Hm$ forms, and the fraction of $\Hm$ ions that survive 
to form $\mHt$. The latter quantity is determined by competition between 
$\mHt$ formation via reaction~\ref{h2f2} and $\Hm$ destruction by mutual 
neutralization with $\Hp$ ions  
 \begin{equation}
 \label{hmneut}
  \Hm + \Hp \rightarrow 2 \mH,
 \end{equation}
and by photodetachment by the incident radiation field
 \begin{equation}
 \label{hmphoto} 
  \Hm + \gamma \rightarrow \mH + \me.
 \end{equation}
Various other reactions also destroy $\Hm$, but these are either significantly 
slower than those above, or become important only at high temperatures, in 
which case any $\mHt$ that does form will very quickly be collisionally 
dissociated. For more details, the reader is referred to the recent reviews
of \citet{aanz}, \citet{gp}, \citet{sld} and \citet{lsd}.

If we assume, for simplicity, that $\Hm$ has already reached its equilibrium
abundance, then we can write the $\mHt$ formation rate as 
\begin{equation}
 \label{general}
 R_{\mHt, \Hm} = k_{1} n_{\me} n_{\mH} \frac{k_{2} n_{\mH}}{k_{2} n_{\mH} +
 k_{5} n_{\Hp} + k_{6}} 
\end{equation} 
where $n_{i}$ is the number density of species $i$, and where the 
rate coefficients $k_{i}$ for the various reactions are listed in 
table~\ref{chem}.

If $\mHt$ formation via reaction~\ref{h2f2} occurs much faster than
the destruction of $\Hm$ by the other reactions, then this reduces to
\begin{equation}
\label{low_x}
 R_{\mHt, \Hm} \simeq k_{1} n_{\me} n_{\mH};
\end{equation} 
in other words, the $\mHt$ formation rate is approximately the same as
the $\Hm$ formation rate.

If, on the other hand, mutual neutralization dominates over $\mHt$ formation
or photodetachment as a means of removing $\Hm$, then equation~\ref{general} 
becomes
\begin{equation}
 R_{\mHt, \Hm} \simeq k_{1} n_{\me} n_{\mH} \frac{k_{2}}{k_{5} x}, 
\end{equation} 
where $x = n_{\Hp}/n_{\mH}$ is the fractional ionization of hydrogen. 
As long as $n_{\me} \simeq n_{\Hp}$, this equation can be further simplified 
to
\begin{equation}
 \label{high_x}
 R_{\mHt, \Hm} \simeq \frac{k_{1} k_{2}}{k_{5}} n_{\mH}^{2}.
\end{equation}

Comparing this equation with equation~\ref{low_x}, we see that for a small 
fractional ionization $R_{\mHt} \propto x$, but that once the fractional
ionization becomes large enough that mutual neutralization dominates, 
$R_{\mHt}$ becomes independent of the ionization: although increases in $x$
still increase the $\Hm$ formation rate, this is balanced by the increase in
the mutual neutralization rate and consequent decrease in the fraction of
$\Hm$ ions surviving to form $\mHt$. This change in behaviour occurs for
fractional ionizations near a critical value $x_{\rm cr}$, defined by
\begin{equation}
\label{xcr}
 x_{\rm cr} = \frac{k_{2}}{k_{5}}.
\end{equation}
The precise value of $x_{\rm cr}$ is somewhat uncertain, due to the 
significant uncertainty that remains in the determination of the mutual
neutralization rate. In this paper, I have chosen to adopt the rate 
listed in \citet{gp}, which is derived from the data of \citet{map}.
This is a conservative choice, in that it gives the lowest value of 
$x_{\rm cr}$; other possibilities include the rates of \citet{dw},
\citet{dl} and \citet{croft}, with the last-named being preferred by the
most recent compilation \citep{lsd}. For the temperature range of interest,
the \citeauthor{gp} rate gives us a value $x_{\rm cr} \sim 5 \times 10^{-3}$,
with only a slight dependence on temperature. The alternative rates typically
give values of $x_{\rm cr}$ that are factors of a few larger. 

When the destruction rate of $\Hm$ ions is dominated by photodetachment,
we obtain another limiting case of equation~\ref{general}
\begin{equation}
 R_{\mHt, \Hm} \simeq k_{1} n_{\me} n_{\mH} \frac{k_{2} n_{\mH}}{k_{6}}.
\end{equation}
This can be written as the $\Hm$ formation rate divided by a suppression 
factor $f_{\rm rad}$:
\begin{equation}
 R_{\mHt, \Hm} \simeq \frac{k_{1} n_{\me} n_{\mH}}{f_{\rm rad}},
\end{equation} 
where
\begin{equation}
 \label{frad} 
f_{\rm rad} = \frac{k_{6}}{k_{2} n_{\mH}}.
\end{equation} 

Determination of the photodetachment rate, and hence $f_{\rm rad}$, requires
knowledge of the incident radiation field. Provided that the opacity of the
gas is low, we can write the photodetachment rate as
\begin{equation}
 k_{6} = 4\pi \int_{\nu_{\rm th}}^{\infty} \frac{\sigma_{\nu} 
J_{\nu}}{h \nu} \, {\rm d\nu}, \label{hmpd} 
\end{equation} 
where $J_{\nu}$ is the mean specific intensity, $\sigma_{\nu}$ is the 
photodetachment cross-section \citep{dj} 
\begin{equation}
 \sigma_{\nu} = 7.928 \times 10^{5} \frac{(\nu - \nu_{\rm th})^{3/2}}{\nu^{3}}
 \: \rm{cm}^{-2}, 
\end{equation} 
and where $h\nu_{\rm th} = 0.755 \: \rm{eV}$ is the energy threshold
for $\Hm$ photodetachment.

Evaluating equation~\ref{hmpd} for the local interstellar radiation field, as
estimated by \citet{mathis}, gives
\begin{equation}
 k_{6} = 4.2 \times 10^{-7} \: \rm{s}^{-1},
\end{equation} 
and so, locally,  
\begin{equation}
 \label{fr}
 f_{\rm rad} = 3.2 \times 10^{2} n_{\mH}^{-1}.
\end{equation} 

If the opacity of the gas is high, then this will overestimate the effects
of radiation. However, at frequencies near the $\Hm$ photodetachment threshold,
the continuum opacity of interstellar gas is low and absorption is dominated
by dust. Consequently, a high opacity implies a high dust content, in which 
case grain-catalyzed formation will dominate.

These limiting cases provide useful insight into the physics of gas-phase
$\mHt$ formation, but in general we must use the full form of 
equation~\ref{general}, which we can rewrite as 
\begin{equation}
 \label{final_hm}
 R_{\mHt, \Hm} = \frac{k_{1} n_{\me} n_{\mH} }{1 + x/x_{\rm cr} 
 + f_{\rm rad}}. 
\end{equation} 

If we now turn to $\mHt$ formation via the $\mHtp$ ion, we find that the 
basic chemistry is remarkably similar. $\mHtp$ is created by the radiative 
association of $\mH$ and $\Hp$ (reaction~\ref{h2f3}), and destroyed by
$\mHt$ formation (reaction~\ref{h2f4}), dissociative recombination 
\begin{equation}
 \mHtp + \me \rightarrow 2\mH,
\end{equation} 
and photodissociation
\begin{equation}
 \mHtp + \gamma \rightarrow \mH + \Hp.
\end{equation} 

As with $\Hm$, the formation of the molecular ion is the limiting step,
with subsequent reactions occurring orders of magnitude faster. If we
again assume that the $\mHtp$ abundance has reached equilibrium, 
then we can write the $\mHt$ formation rate as
\begin{equation}
 R_{\mHt, \mHtp} = k_{3} n_{\Hp} n_{\mH} \frac{k_{4} n_{\mH}}{k_{4} n_{\mH} 
+ k_{7} n_{\me} + k_{8}},
\end{equation} 
which has the same form as equation~\ref{general}. Indeed, we can
rewrite it as
\begin{equation}
 \label{final_h2p}
 R_{\mHt, \mHtp} = \frac{k_{3} n_{\Hp} n_{\mH} }{1 + x/x_{\rm cr} 
+ f_{\rm rad}}, 
\end{equation} 
only now
\begin{equation}
\label{xcr_h2p}
 x_{\rm cr} = \frac{k_{4}}{k_{7}}, 
\end{equation}
and 
\begin{equation}
\label{frad_h2p} 
f_{\rm rad} = \frac{k_{8}}{k_{4} n_{\mH}},
\end{equation} 
where the photodissociation rate, $k_{8}$, is calculated in a similar fashion
to the photodetachment rate above. Evaluating these, we find that $x_{\rm cr}$
for $\mHtp$ is typically an order of magnitude larger than for $\Hm$, and that
in the local ISM
\begin{equation}
 f_{\rm rad} \simeq n_{\mH}^{-1},
\end{equation}
where I have again used the \citet{mathis} radiation field, together with the
$\mHtp$ photodissociation cross-section from \citet{sk}.

Comparing these values with those for $\Hm$, we see that $\mHtp$ is 
significantly more robust. However, it forms at a much slower rate 
(between two and three orders of magnitude, depending on temperature) and 
so in most cases $\Hm$ dominates. Nevertheless, there are exceptions, as
we will see in section~\ref{astro}.

Finally, a few other possible mechanisms have been suggested for gas-phase
$\mHt$ formation. \citet{lb} propose that $\mHt$ can form as a result of
direct radiative association
\begin{equation}
 \mH + \mH \rightarrow \mHt + \gamma,
\end{equation} 
provided that one of the hydrogen atoms is in an excited electronic state.
\citet{rdb} show that a more efficient mechanism is formation
of $\mHtp$ by associative ionization
\begin{equation}
 \mH + \mH \rightarrow \mHtp + \me
\end{equation} 
with the $\mHtp$ thereafter forming $\mHt$ by reaction~\ref{h2f4} above. 
This mechanism again requires one of the hydrogen atoms to be in an excited
atomic state. However, this requirement means that in general these
reactions are not important, as the necessary population of excited
atomic hydrogen is only found in a few unusual circumstances, such as in 
the universe immediately after recombination.

\subsection{Grain-catalyzed $\mHt$ formation} 
\label{grain}
Despite its importance in local interstellar chemistry, the rate of 
$\mHt$ formation on dust grains is still uncertain. In local molecular 
clouds, observations suggest a formation rate \citep{jura}
\begin{equation}
 k_{9} \sim 3 \times 10^{-17} \: \rm{cm}^{3} \: \rm{s}^{-1}. 
\end{equation}
Observations of $\mHt$ in the  LMC and SMC with the {\sc fuse} satellite
\citep{tum} suggest a value that is an order of magnitude smaller, but
this is consistent with the underlying rate per unit dust mass being the
same, since the mean dust-to-gas ratio within these galaxies is significantly
smaller than in the Milky Way \citep{issa}.

Unfortunately, direct measurements of this kind can only give us information
about $\mHt$ formation in physical conditions that are easily accessible to
observations, and provide little basis on which to predict the $\mHt$ 
formation rate in different regimes. For this, we must turn to theory. 

A large body of theoretical work exists on the subject of $\mHt$ formation
on grains \citep[see, for example, the review of][and the many references 
therein]{pir}, stretching back almost forty years to the pioneering work of
\citet{gs}. In a highly influential paper, \citet{hws} parameterized 
the $\mHt$ formation rate as
\begin{equation}
 R = 0.5 \bar{v}_{\mH} \sigma_{\rm d} S f_{a} n_{\mH} n_{\rm d},  
\end{equation} 
where $n_{\rm d}$ is the number density of dust grains, $\sigma_{\rm d}$
is their mean geometric cross-section, $\bar{v}_{\mH}$ is the mean
velocity of the hydrogen atoms striking the grains, $S$ is the sticking 
coefficient (the probability that a hydrogen atom striking the grain will 
stick to the surface) and $f_{a}$ is the fraction of adsorbed hydrogen atoms 
that actually form $\mHt$, rather than simply escaping back into the gas phase.
They argued that for gas and grain temperatures typical of molecular clouds, 
both $S$ and $f_{a}$ should be of order unity. 

\citet{hm} later used this prescription to derive an $\mHt$ formation rate
for the local ISM that continues to be widely cited:
\begin{equation}
 \label{dust}
 k_{9} = 3 \times 10^{-17} \frac{T_{2}^{0.5} f_{a}}{y} \: \rm{cm}^{3}
 \: \rm{s}^{-1},
\end{equation}
where
\begin{equation}
y = 1 + 0.4 \sqrt{T_{2} + T_{{\rm gr},2}} + 0.2T_{2} + 0.08T_{2}^{2}, 
\end{equation}
and where $T_{2}$ and $T_{{\rm gr},2}$ are the gas and grain temperatures in
units of $100 \: \rm{K}$. They argued that $f_{a}$ should be approximately 
constant and of order unity for grain temperatures below some critical value 
$T_{\rm cr}$, but that for $T_{\rm gr} > T_{\rm cr}$, it should fall off 
exponentially, with most of the hydrogen atoms evaporating from the grain 
surface before they have time to form $\mHt$. The value of $T_{\rm cr}$ has 
proved hard to determine precisely, but is of the order of $100 \: \rm{K}$.

Although this rate has been widely adopted in the literature, recent 
experiments have cast doubt on its validity at high temperatures, and
suggest that the $\mHt$ formation rate may be smaller than previously assumed
(Pirronello {\em et~al.} 1997a,b, 1999; Katz {\em et al.} 1999; 
Biham {\em et~al.} 2001). However, since this conclusion is not entirely 
clear \citep{tiel} and their work is still ongoing, I have tentatively 
adopted the \citeauthor{hm} rate below, with the proviso that the values
of ${\cal D}_{\rm cr}$ that I derive may prove to be lower limits if the 
results of Pirronello {\em et~al.} are borne out by future work.

\section{Comparing the different modes of formation}
\label{comp}
We can combine equations~\ref{final_hm} and \ref{final_h2p} to write
the total gas-phase $\mHt$ formation rate as
\begin{equation}
 R_{\mHt, {\rm gas}} = R_{\mHt, \Hm} + R_{\mHt, \mHtp}, 
\end{equation}
while the grain-catalyzed rate can be written as
\begin{equation}
 \label{simple2} 
 R_{\mHt, {\rm dust}} =  k_{9} n_{\rm tot} n_{\mH} 
\left( \frac{\cal D}{\cal D_{\rm MW}} \right),
\end{equation} 
where $n_{\rm tot}$ is the total particle number density, ${\cal D}$ is the 
dust-to-gas ratio and ${\cal D_{\rm MW}}$ is its value in the local ISM.

Combining these equations, we can easily solve for the dust-to-gas ratio
at which $R_{\mHt, {\rm gas}}$ and $ R_{\mHt, {\rm dust}}$, which I
denote as ${\cal D}_{\rm cr}$: 
\begin{equation}
 \label{result} 
 {\cal D_{\rm cr}} = \frac{R_{\mHt, {\rm gas}}}{k_{9} n_{\rm tot} n_{\mH}}
 \: {\cal D_{\rm MW}}. 
\end{equation} 
In the common case that $R_{\mHt, \Hm} \gg R_{\mHt, \mHtp}$, this equation
reduces to  
\begin{equation}
 \label{result2} 
 {\cal D_{\rm cr}} = \frac{k_{1}}{k_{9}} \frac{x}{1 + x/x_{\rm cr} 
+ f_{\rm rad}} \: {\cal D_{\rm MW}}, 
\end{equation} 
where $x_{\rm cr}$ and $f_{\rm rad}$ are given by equations~\ref{xcr} and 
\ref{frad} respectively. A similar equation can be written in the much 
less common case that $R_{\mHt, \mHtp} \gg R_{\mHt, \Hm}$.

\begin{figure}
\epsscale{0.8}
\plotone{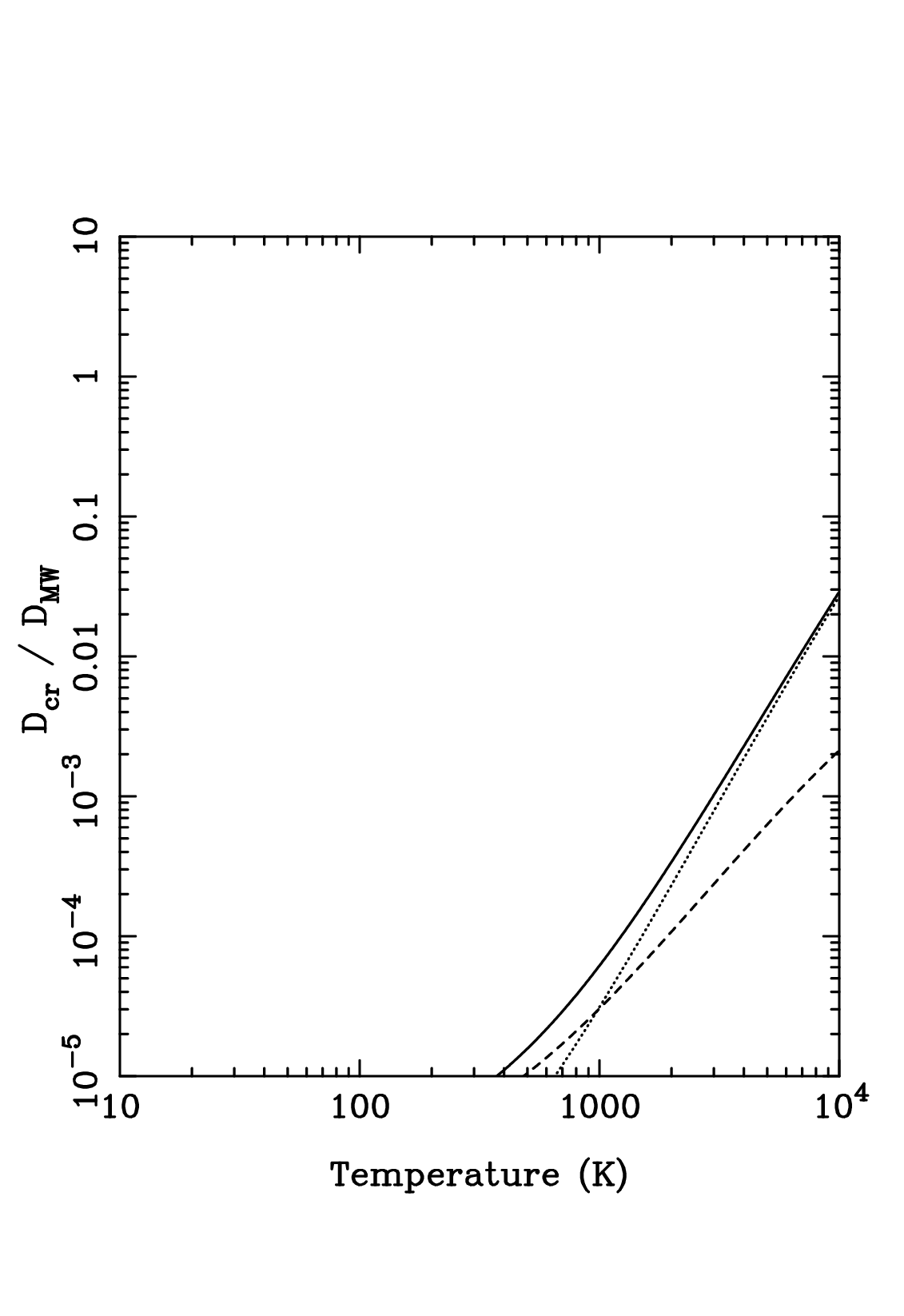} 
\caption{ ${\cal D}_{\rm cr}$ as a function of temperature for low ionization,
low density gas ($x = 10^{-4}, \: n_{\mH} = 1 \: \rm{cm}^{-3}$). 
The solid line is the value of ${\cal D}_{\rm cr}$ given by 
equation~\ref{result}; the dashed and dotted lines are the contributions 
to this value of $\Hm$ and $\mHtp$ respectively. \label{fig1}}
\end{figure}

\begin{figure}
\epsscale{0.8}
\plotone{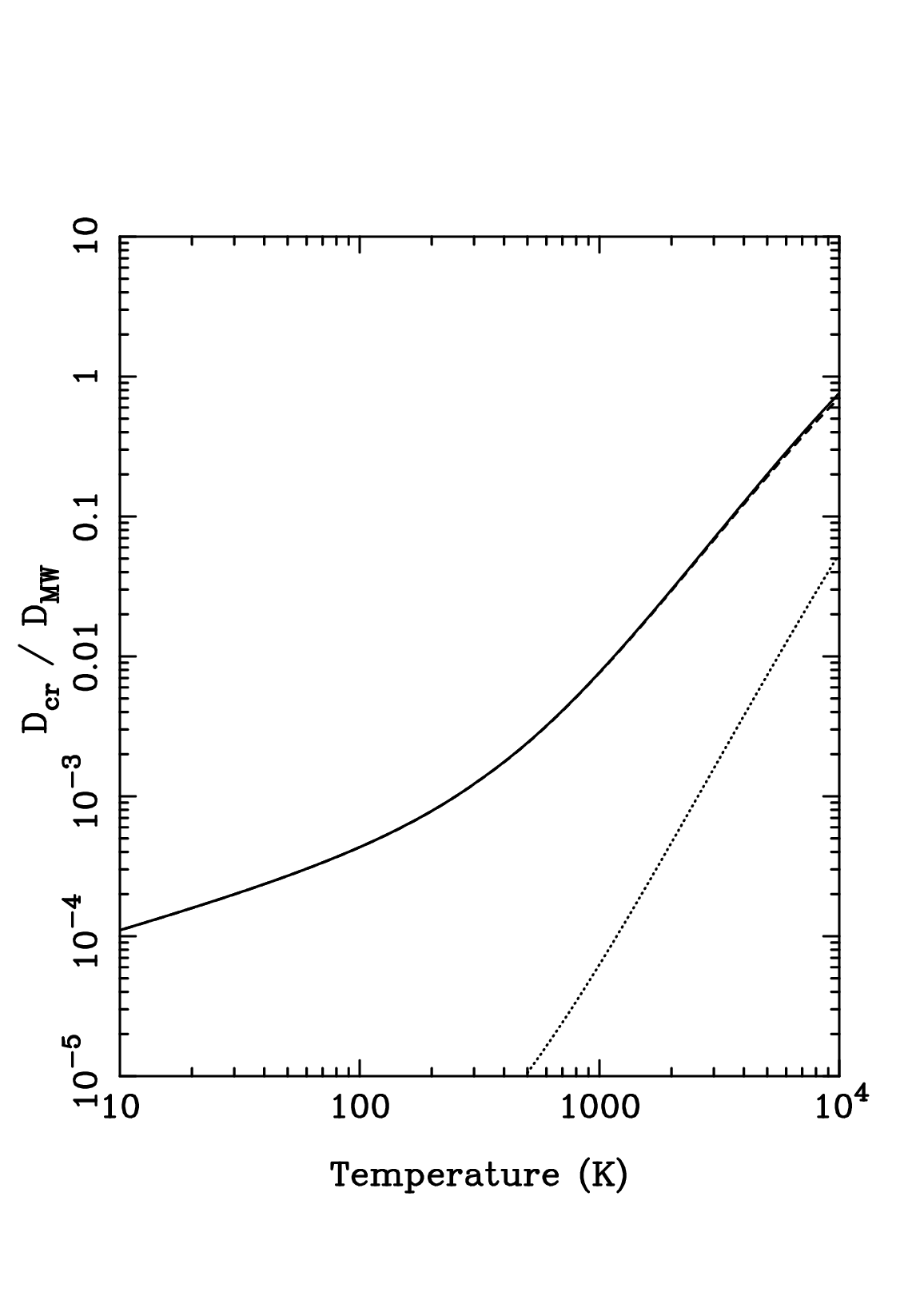}
\caption{As figure~\ref{fig1}, but for low ionization, high density gas 
($x = 10^{-4}, \: n_{\mH} = 10^{3} \: \rm{cm}^{-3}$). \label{fig2}}
\end{figure}

\begin{figure}
\epsscale{0.8}
\plotone{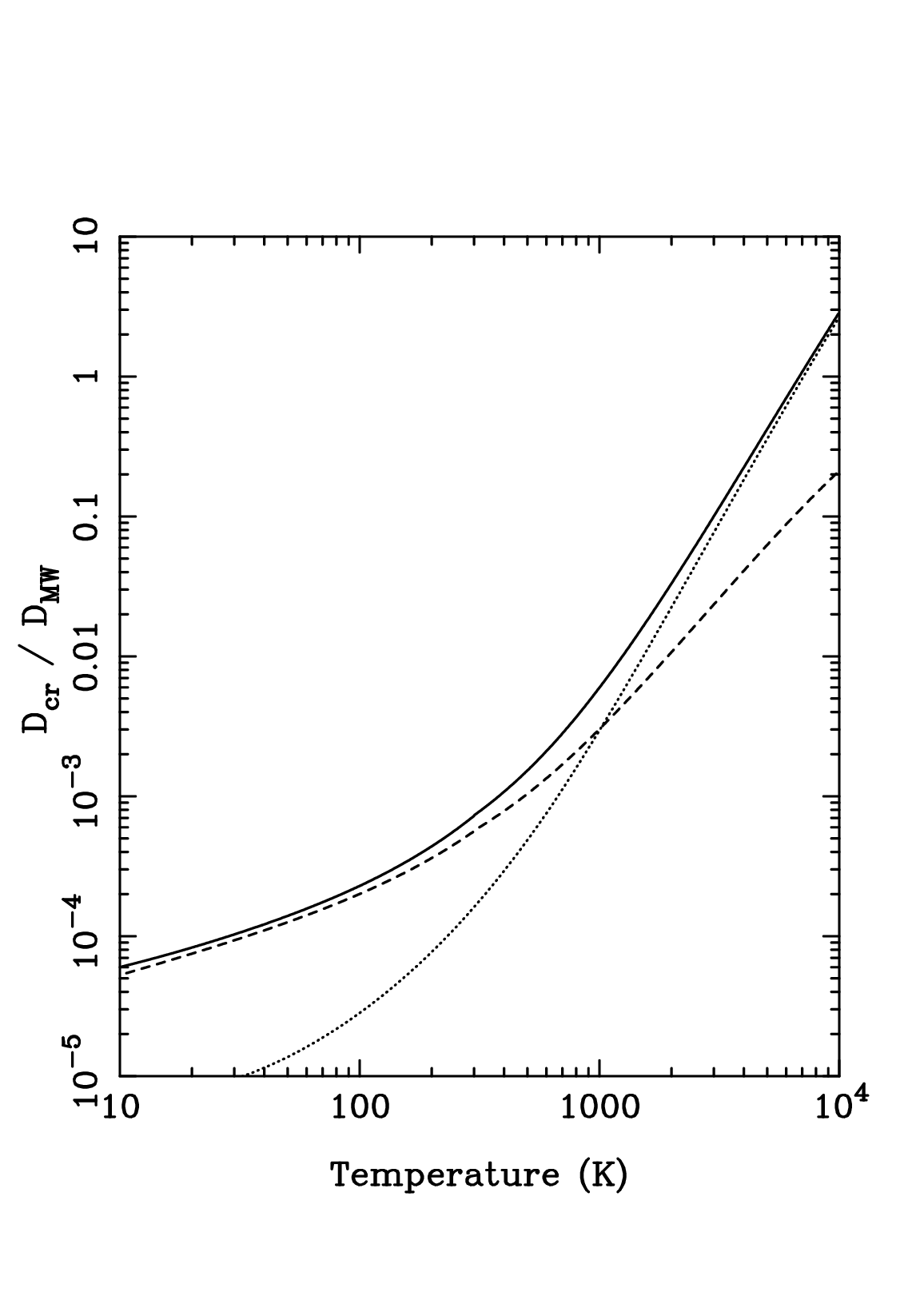}
\caption{As figure~\ref{fig1}, but for high ionization, low density gas 
($x=10^{-2}, \: n_{\mH} = 1 \: \rm{cm}^{-3}$).
\label{fig3}}
\end{figure}

\begin{figure}
\epsscale{0.8}
\plotone{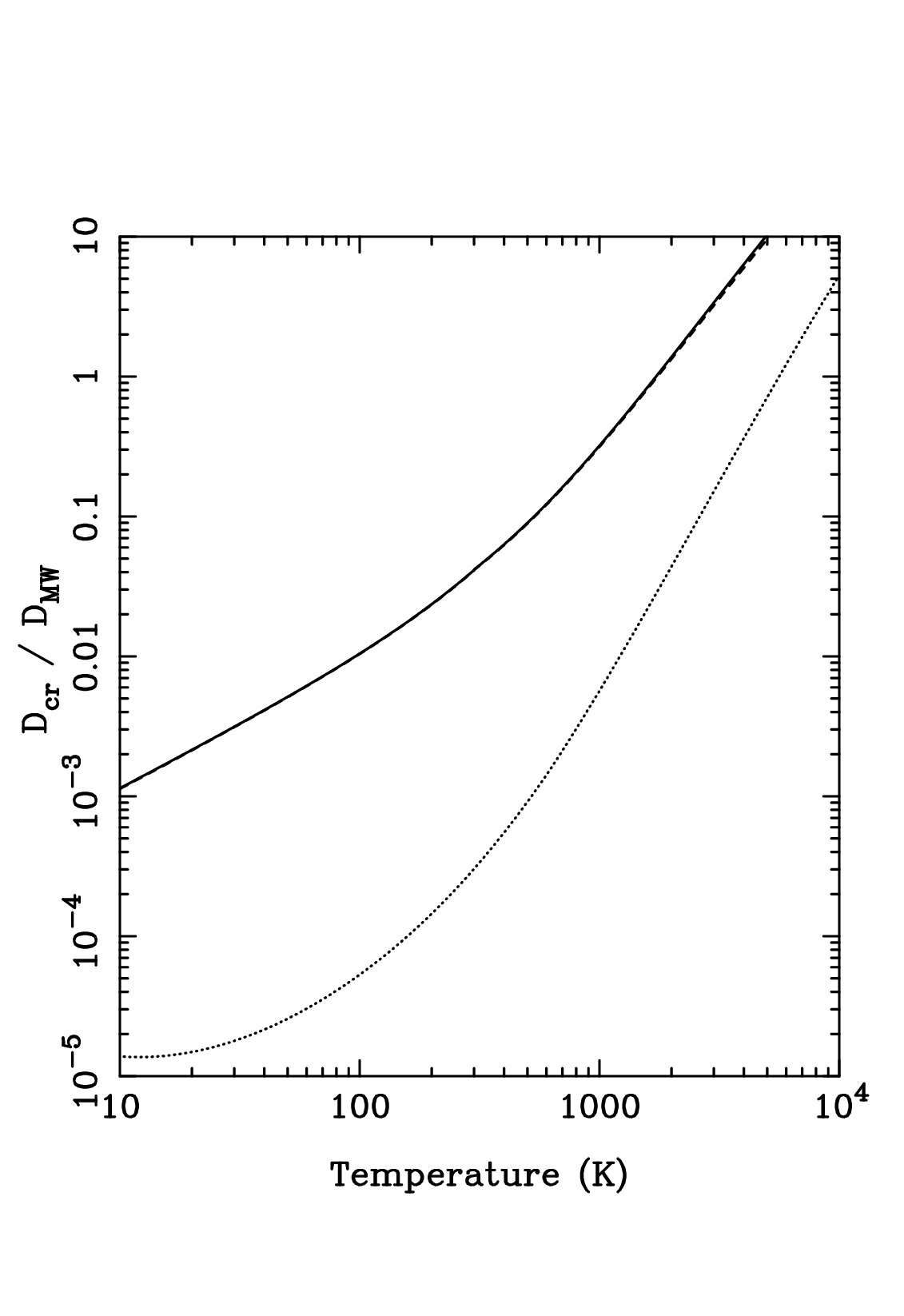}
\caption{As figure~\ref{fig1}, but for high ionization, high density gas 
($x=10^{-2}, \: n_{\mH} = 10^{3} \: \rm{cm}^{-3}$).
\label{fig4}}
\end{figure}

In order to help illustrate the behaviour of these equations, I plot in 
figures~\ref{fig1} to \ref{fig4} the value of ${\cal D}_{\rm cr}$ as a 
function of temperature for gas illuminated by the \citet{mathis} radiation
field in four different scenarios: low ionization, low density gas 
($x = 10^{-4}, n_{\mH} = 1 \: \rm{cm}^{-3}$; figure~\ref{fig1}), low 
ionization, high density gas ($x = 10^{-4}, n_{\mH} = 10^{3} \: \rm{cm}^{-3}$; 
figure~\ref{fig2}), high ionization, low density gas ($x = 10^{-2}, 
n_{\mH} = 1 \: 
\rm{cm}^{-3}$; figure~\ref{fig3}) and high ionization, high density gas 
($x = 10^{-2}, n_{\mH} = 10^{3} \: \rm{cm}^{-3}$; figure~\ref{fig4}).
In each case, I adopt a fixed grain temperature $T_{\rm gr} =20 \: \rm{K}$,
although small changes in $T_{\rm gr}$ have little effect on the results
provided that it remains less than $T_{\rm cr}$.  

A striking feature of these plots is the strong temperature dependence
of ${\cal D}_{\rm cr}$. At low temperatures, grain-catalyzed $\mHt$ formation
is relatively efficient and very little dust is needed before grain catalysis
dominates. Above a few hundred K, however, the efficiency of grain 
catalysis decreases significantly, while the efficiency of gas-phase
$\mHt$ formation continues to grow. As a result, the required dust abundance
rises sharply with increasing temperature.  

\section{Astrophysical examples}
\label{astro}
From the behaviour outlined in figures~\ref{fig1}--\ref{fig4}, it is clear
that gas-phase $\mHt$ formation is at its most effective in warm, dense gas
with a high fractional ionization. However, most of the molecular gas 
that we observe in our galaxy is in the form of molecular clouds with 
low temperatures ($T \sim 20 \: \rm{K}$) and very low fractional ionizations 
($x \sim 10^{-7}$) and in these conditions grain catalyzed formation
dominates by many orders of magnitude.

A more promising place to look for gas-phase $\mHt$ formation is in the
so-called warm neutral medium (WNM). In models of the multiphase ISM that
assume thermal pressure equilibrium between phases \citep{fgh,mo,wolf},
this is predicted to have a temperature of approximately $8000 \: \rm{K}$, 
high enough to collisionally dissociate $\mHt$. However, recent observations 
\citep{ht2} and simulations that include the effects of turbulence 
\citep{gaz,mac} suggest that much of this gas is actually at much lower
temperatures; for instance, \citeauthor{ht2} quote a temperature range of
$500 < T < 5000 \: \rm{K}$. 

Taking representative values for the temperature and ionization of the WNM 
to be $T = 2000 \: \rm{K}$ and $x = 10^{-2}$ \citep{hei}, I find that
\begin{equation}
 \label{WNM} 
 \frac{{\cal D_{\rm cr}}}{{\cal D_{\rm MW}}} =  
 \frac{4.1}{2.7 + f_{\rm rad, \Hm}} 
                   + \frac{0.047}{1.1 + f_{\rm rad, \mHtp}}.
\end{equation} 
In other words, gas-phase formation would dominate if we could ignore the
effects of the radiation field. In practice, this is not possible; at the 
densities characteristic of the WNM ($n \simeq 0.1 \: \rm{cm}^{-3}$),
we have
\begin{eqnarray}
f_{\rm rad, \Hm} & = & 3.9 \times 10^{3} \\
f_{\rm rad, \mHtp} & = & 1.0 \times 10^{2}, 
\end{eqnarray}
and equation~\ref{WNM} becomes
\begin{equation}
 {\cal D_{\rm cr}} = 5.2 \times 10^{-3} \: {\cal D_{\rm MW}}.
\end{equation}

These two examples demonstrate that gas-phase $\mHt$ formation is unimportant
in the bulk of the gas in the Milky Way: either the temperature and ionization
are too low, as in molecular clouds, or the gas is too diffuse and $\mHt$ 
formation is suppressed by the photodissociation of $\Hm$ and $\mHtp$.

However, there are a few counterexamples. For instance, gas-phase $\mHt$ 
formation has long been known to play an important role in the chemistry
of nova ejecta \citep{raw} and protostellar outflows \citep{gmh}, where
the gas initially has little or no dust (although more generally forms later)
and where the high gas densities help mitigate the effects of 
photodissociation. Gas-phase formation is also predicted to dominate
the molecular chemistry of freely-expanding supernova remnants such as
SN1987A \citep{cm}. Finally, \citet{lmc} suggest that gas-phase 
formation may dominate in X-ray dissociation regions (dense clouds 
illuminated by hard X-rays); modelling by \citet{mht} would appear to confirm 
this.

These are somewhat unusual conditions, however, and in general dust 
abundances significantly below the typical galactic value are required 
before gas-phase $\mHt$ formation becomes competitive with grain-catalyzed 
formation. 

One place in which we might expect to find these low dust abundances is
in the metal-poor gas within dwarf galaxies. \citet{kh} examine a well-studied
example, the metal-poor dwarf IZw18, and show that provided that its neutral
ISM is clumpy (with clumps densities $n \gtrsim 100 \: \rm{cm}^{-3}$) 
and  moderately ionized ($x \sim 10^{-3}$), then gas-phase formation will 
dominate. They also show that the formation of $\mHt$ in this manner would 
not conflict with the upper limit on the $\mHt$ column density of IZw18 
obtained by {\sc fuse} \citep{fuse}, due to the small filling factor of the
clumps.

Another place we might look for significant gas-phase $\mHt$ formation is
in damped Lyman-$\alpha$ (DLA) systems, many of which have low dust abundances 
\citep[see, eg][]{lopez}. An interesting example is the absorber at 
$z = 3.025$ in the spectrum of Q0347-3819 recently studied by \citet{lev}. 
This system has a temperature $T \simeq 400 \: \rm{K}$ (as inferred from the
Doppler broadening of its many associated $\mHt$ and metal absorption lines),
and a fractional ionization $x \simeq 2 \times 10^{-5}$. If we assume that
$\Hm$ photodetachment is negligible, we find that for this system
\begin{equation}
 D_{\rm cr} = 4.6 \times 10^{-4} \: {\cal D}_{\rm MW}.
\end{equation} 
Comparing this with a measured dust-to-gas ratio of ${\cal D} \simeq 0.05 
{\cal D}_{\rm MW}$, we see that gas-phase $\mHt$ formation contributes no
more than about 1\% of the total $\mHt$ in this system. Including the effects
of radiation merely strengthens this conclusion. 

There is no reason to suspect that this situation is particularly unusual;
all damped Lyman-$\alpha$ systems {\em by definition} have large \hi column 
densities, and consequently will have small fractional ionizations. We would
therefore expect grain-catalysis to dominate in these systems.

Finally, gas-phase $\mHt$ formation has long been known to play an important
role in the early stages of galaxy formation. In primordial gas, this is 
obvious: there is no dust, so any $\mHt$ that forms {\em must} form in the 
gas phase. A more interesting problem is determining the value of 
${\cal D}_{\rm cr}$ for these systems; in other words, at what point does
grain-catalyzed formation overtake gas-phase formation? 

For the purposes of this discussion, I adopt the example of an $\mHt$-cooled
protogalaxy with temperature $T = 1000 \: \rm{K}$ and fractional ionization
$x = 2 \times 10^{-4}$ \citep{teg}. These values are appropriate for the
first generation of star-forming protogalaxies, and while they may be 
underestimates for later generations, my analysis can easily be rescaled
for higher values. For this example protogalaxy, I find that
\begin{equation}
 \label{my_pg}
 \frac{{\cal D_{\rm cr}}}{{\cal D_{\rm MW}}} =  
 \frac{2.1 \times 10^{-2}}{1 + f_{\rm rad, \Hm}} 
                   + \frac{1.3 \times 10^{-4}}{1 + f_{\rm rad, \mHtp}}.
\end{equation} 
Thus, if the radiation field is unimportant, ${\cal D}_{\rm cr} \simeq
0.02 \, {\cal D}_{\rm MW}$, comparable to the values seen in some 
metal-poor dwarf galaxies at the present-day \citep{lf}.

How strong is the radiation field within a protogalaxy? There are potentially
three main contributors to this field: the protogalaxy's own stellar 
population, emission from neighbouring galaxies and radiation from the 
cosmological background produced by distant sources. It is simplest to
consider these separately.

Much of the optical and near-infrared radiation responsible for destroying 
$\Hm$ and $\mHtp$ is produced by long-lived stars, and so the contribution
of the protogalaxy's stellar population depends as much on its star formation 
history as on its current star formation rate. This makes it very difficult
to parameterize its effects in the general case; it is much easier to examine
a simple example that will hopefully be broadly representative.

For the purposes of this example, I assume:
\begin{enumerate}
\item That the protogalaxy underwent an instantaneous (or near-instantaneous)
starburst $10^{8} \: \rm{yr}$ ago, following which it has formed no more 
stars.

\item That the stars which did form are located in the centre of the 
protogalaxy, within a small enough region that I can approximate their 
emission as coming from a point source.

\item That the luminosity and spectral energy distribution of this stellar
cluster are well described by the $Z = 0.05 Z_{\odot}$ model of \citet{lei}.

\item That the protogalaxy itself is well-described by a truncated isothermal
sphere density profile \citep{iliev}.
\end{enumerate} 
All of these assumptions are debatable, but they do provide us a basis on 
which to estimate the effects of the stellar radiation field. Moreover,
these assumptions are somewhat conservative, and tend to minimize the 
effectiveness of the stellar radiation. For instance, if we reduce the
time since the starburst from $10^{8} \: \rm{yr}$ to $10^{7} \: \rm{yr}$,
then the photodetachment rate increases by a factor of fifty. Similarly,
if we assume continuous star-formation rather than an instantaneous 
starburst, then we obtain a similar (or slightly larger) photodetachment 
rate once the total mass of stars formed has reached a comparable level.

We could also criticize the adoption of the \citeauthor{lei} model, on the
basis that it assumes a standard Salpeter IMF, while there is considerable
evidence that the the primordial IMF is biased towards high masses 
\citep{lar}. However, this again means that we will underestimate the
photodetachment rate (although we will significantly overestimate the 
lifetime of the stellar population).

Returning to my example, the first two assumptions allow me to write 
the $\Hm$ photodetachment rate at a distance $R$ from the stars as
\begin{equation}
 k_{6} = \frac{1}{R^{2}} \int_{\nu_{\rm th}}^{\nu_{0}} \frac{\sigma_{\nu} 
L_{\nu}}{h \nu} \, {\rm d\nu},
\end{equation} 
where $L_{\nu}$, the stellar luminosity per unit frequency, is given
by the \citeauthor{lei} model. Using this value, I obtain
\begin{equation}
 k_{6} = 4.9 \times 10^{-8} F(M_{*}, R) \: \rm{s}^{-1},
\end{equation}
where
\begin{equation}
 F(M_{*}, R) =  \left( \frac{M_{*}}{10^{6} \: 
\rm{M}_{\odot}} \right) \left( \frac{R}{1 \: \rm{kpc}} \right)^{-2},
\end{equation} 
and where $M_{*}$ is the mass of stars formed in the starburst. Similarly,
we can write the $\mHtp$ photodissociation rate as
\begin{equation}
 k_{8} = 9.4 \times 10^{-10}  F(M_{*}, R) \: \rm{s}^{-1},
\end{equation} 
and from these rates calculate $f_{\rm rad, \Hm}$ and $f_{\rm rad, \mHtp}$
\begin{eqnarray}
 f_{\rm rad, \Hm} & = &   40 \, n_{\mH}^{-1}  F(M_{*}, R) \label{first_fr} \\
 f_{\rm rad, \mHtp} & = & 1.5 \, n_{\mH}^{-1}  F(M_{*}, R). \label{second_fr}
\end{eqnarray} 
By comparing these values and equation~\ref{my_pg}, we can see that 
formation via $\mHtp$ contributes at most about 10\% of the $\mHt$ 
produced in the gas phase, with the rest coming from $\Hm$.

To evaluate these numbers, I use the fact that for a truncated isothermal 
sphere, 
\begin{equation}
 n_{\mH} \left( \frac{R}{1 \: \rm{kpc}} \right)^{2} \simeq 6.9 \times 10^{-3}
 \frac{\Omega_{b}}{\Omega_{m}}
\end{equation}
in regions outside of the core. The final unknown, $M_{*}$, can be written as
\begin{equation} 
 M_{*} = 1.5 \times 10^{7} \varepsilon_{*} (1+z)^{-3/2} 
 \left( \frac{\Omega_{b}}{\Omega_{m}^{3/2} h} \right) \: {\rm M}_{\odot}, 
\label{last_mass} 
\end{equation} 
where $\varepsilon_{*}$ is the star formation efficiency of the protogalaxy, 
$z$ is its redshift of formation and $h$ is the Hubble constant in units of 
$100 \: \rm{km} \: \rm{s}^{-1} \: \rm{Mpc}^{-1}$.

For a protogalaxy that formed in a standard $\Lambda$CDM cosmology 
($\Omega_{m} = 0.3$, $\Omega_{b} = 0.04$, $h = 0.7$) at a redshift $z=10$,
and that formed stars with an efficiency $\varepsilon_{*} = 0.01$, we
find that
\begin{equation}
 \label{int_rad}
 {\cal D}_{\rm cr} \simeq 3.8 \times 10^{-4} {\cal D}_{\rm MW},
\end{equation} 
Thus, in this particular example, radiation from the existing stellar 
population reduces ${\cal D}_{\rm cr}$ by almost two orders of magnitude. 

In view of the uncertainties involved in producing this estimate, it
would be unwise to over-generalize. However, since my assumptions verge
on the conservative side, it seems likely that in realistic protogalactic
models we would see similar effects, and that gas-phase $\mHt$ formation 
will rapidly be overtaken by grain-catalyzed formation. 

What about protogalaxies that have yet to form stars? In this case, there
is no significant local contribution to the radiation field, which instead
is produced by neighbouring sources and/or the cosmological background. 

For neighbouring sources, we can reuse the above formalism, as long as we
set $R$ to the distance to the extragalactic source. However, this is 
typically an order of magnitude or more greater than the size of a 
protogalaxy, implying that the effect of the radiation will be {\em at least}
two orders of magnitude smaller than the effects discussed above. 
Consequently, radiation from protogalaxies of the size discussed here will 
have little or no effect on gas-phase $\mHt$ formation within their 
neighbours, unless their emitted flux is substantially larger than has
been assumed here. 

For the background, we again face the problem that any conclusions that we 
can draw are strictly limited by our poor knowledge of the star formation 
history, this time on a cosmological rather than protogalactic scale. The 
best that we can do is to determine how strong the background needs to be
before it has a significant effect. Modeling the background below the 
Lyman limit as a power-law,
\begin{equation}
 J_{\nu} = J_{21} \left( \frac{\nu_{0}}{\nu} \right)^{\alpha},  
\end{equation} 
where $J_{21} = 10^{-21} \: \rm{erg} \: \rm{s}^{-1} \: \rm{cm}^{-2} \: 
\rm{Hz}^{-1} \: \rm{sr}^{-1}$ and where $h \nu_{0} = 13.6 \: \rm{eV}$,
I find that for $\alpha = 1$, 
\begin{equation} 
 k_{6} = 8.1 \times 10^{-10} J_{21} \: \rm{s}^{-1},
\end{equation} 
and hence
\begin{equation} 
 f_{\rm rad, \Hm} = 0.6 \, J_{21} \, n_{\mH}^{-1}. 
\end{equation} 

The significance of the background varies with $n_{\mH}$ and hence with
position within the protogalaxy. For my example protogalaxy, formed at a 
redshift $z = 10$, gas near the truncation radius has a density
$n_{\mH} \sim 7 \times 10^{-3} \: \rm{cm}^{-3}$ and thus is affected for
$J_{21} \gtrsim 0.01$; on the other hand, gas in the central core has
$n_{\mH} \sim 1 \: \rm{cm}^{-3}$ and is only affected for $J_{21} \gtrsim 
1$.

\section{Conclusions}
\label{conc}
The simplicity of the basic chemistry involved in gas-phase $\mHt$ formation 
means that it is easy to construct a fairly accurate expression for the 
formation rate in terms of only a few parameters: the temperature, density and
fractional ionization of the gas, plus the strength of the radiation field
near the $\Hm$ and $\mHtp$ photodissociation thresholds. Expressions for
the rate of $\mHt$ formation via the $\Hm$ and $\mHtp$ ions are given by
equations~\ref{final_hm} \& \ref{final_h2p} respectively, and the total 
formation rate is simply the sum of these two values.

Using these expressions, together with an analytical expression for the
grain-catalyzed $\mHt$ formation rate, one can solve for ${\cal D}_{\rm cr}$,
the dust-to-gas ratio required for grain-catalyzed $\mHt$ formation to 
overtake gas-phase formation. The results demonstrate that, in principle,
gas-phase $\mHt$ formation could be comparable to grain-catalyzed formation
in galactic gas, particularly at high temperatures where the latter is 
inefficient. In practice, however, it is usually significantly slower,
either because of a shortage of free electrons and protons (which reduces
the formation rate of the intermediate ions) or because the incident
radiation field destroys the ions before they have a chance to form $\mHt$.

Finally, in order to demonstrate the simplicity and potential utility of 
this approach,
I have applied it to a number of astrophysical examples. Not surprisingly, I 
find that in most cases ${\cal D}_{\rm cr}$ is significantly less than the
mean galactic value, often by several orders of magnitude. Nevertheless, there
are counterexamples, such as X-ray photodissociation regions \citep{lmc,mht},
nova ejecta \citep{raw}, or the high-redshift protogalaxies analyzed in 
detail here. A common thread linking many of these exceptions seems to be 
the fact that they have dust-to-gas ratios (but not necessarily 
metallicities) significantly lower than the mean galactic value. 

\acknowledgments
I would like to acknowledge useful comments on an earlier draft
of this paper from Mordecai Mac-Low and Michael D. Smith. I would also
like to thank the anonymous referee for a timely and useful report, and
for bringing the work of \citeauthor{lmc} to my attention.
This work was supported by NSF grant AST99-85392.

\clearpage

\begin{deluxetable}{llc}
\tablecaption{Reaction rates \label{chem}}
\tablewidth{0pt}
\tablehead{
\colhead{Reaction} & \colhead{Rate (${\rm cm^{-3} \: s^{-1}}$)}  
& \colhead{Reference} }  
\startdata
1. $\mH + \me \rightarrow \Hm + \gamma$ & 
$k_{1} = 1.4 \times 10^{-18} T^{0.928} \exp \left(-\frac{T}{16200} \right)$
& \citet{dj} \\
 & & \\
2. $\Hm + \mH \rightarrow \mHt + \me$ & 
$k_{2} = \left\{ \begin{array}{l} 1.5 \times 10^{-9} \\
4.0 \times 10^{-9} T^{-0.17} 
\end{array} 
\right. $ 
\hfill $ \begin{array}{c}
T < 300 {\rm K} \\
T > 300 {\rm K}
\end{array} $ & \citet{ldz} \\
 & & \\
3. $\mH + \Hp \rightarrow \mHtp + \gamma$ &
$k_{9} = \begin{array}{l} 
 \rm{dex}[-19.38 - 1.523\log T \\
 \mbox{} + 1.118\log(T)^{2} - 0.1269\log(T)^{3}] 
 \end{array}$
& \citet{rp} \\
 & & \\
4. $\mHtp + \mH \rightarrow \mHt + \Hp$ & 
$k_{4} = 6.4 \times 10^{-10}$
& \citet{kah} \\
 & & \\
5. $\Hm + \Hp \rightarrow 2 \mH$ & 
 $ k_{5} = \begin{array}{l} 5.7 \times 10^{-6} T^{-0.5} + 6.3 \times 10^{-8} \\
           \mbox{} -9.2 \times 10^{-11} T^{0.5} +4.4 \times 10^{-13} T 
           \end{array}$  
& \citet{map} \\
& & \\
6. $\Hm + \gamma \rightarrow \mH + \me$ & \multicolumn{1}{c}{{\rm See text}}
& --- \\
& & \\
7. $\mHtp + \me \rightarrow 2\mH$ & $k_{7} = \left\{ 
\begin{array}{l}
  1.0\times 10^{-8}  \\
  1.32 \times 10^{-6} T^{-0.76} 
\end{array} 
\right.$ \hfill $ 
\begin{array}{c}
T < 617 \rm{K} \\
T > 617 \rm{K}
\end{array} $  
& \citet{sdgr} \\
& & \\
8. $\mHtp + \gamma \rightarrow \mH + \Hp$ & \multicolumn{1}{c}{{\rm See text}}
& --- \\
& & \\
9. $\mH + \mH + {\rm grain} \rightarrow \mHt + {\rm grain}$ & 
\multicolumn{1}{c}{{\rm See text}} &
--- \\  
\enddata
\end{deluxetable}

\end{document}